\documentstyle[prb,aps]{revtex}
\begin{document}
\draft
\title{Weak Coupling Phases of the Attractive $t-t'$ Hubbard Model
     at the Van Hove Filling}
\author{J. V. Alvarez $^1$ and J. Gonz\'alez $^2$ \\}
\address{
        $^1$Departamento de Matem\'aticas.
        Universidad Carlos III.
        Butarque 15.
        Legan\'es. 28913 Madrid. Spain. \\
        $^2$Instituto de Estructura de la Materia.
        Consejo Superior de Investigaciones Cient{\'\i}ficas.
        Serrano 123, 28006 Madrid. Spain.}
\date{\today}
\maketitle
\begin{abstract}

We apply a wilsonian renormalization group approach to the
continuum limit of the attractive $t-t'$ Hubbard model, taken when
the Fermi level is at the Van Hove singularity of the density of
states. The model has well-defined scaling properties and
the effective couplings display an unbounded flow in the
infrared. We determine the leading instabilities by computing
the different response functions up to $t' = 0.5 t$. The phase
diagram shows a large boundary between superconducting and
charge-density-wave phases, that merge in a triple point with a
phase separation instability. The latter is realized down to
very low coupling constant, as the Fermi sea degenerates towards
a pair of straight lines near $t' = 0.5 t$.

\end{abstract}

During recent years there has been important progress in
understanding the properties of quantum electron liquids in
dimension $D < 3$. One of the most fruitful approaches in this
respect springs from the use of renormalization group (RG)
methods, in which the different liquids are characterized by
several fixed-points controlling the low-energy properties. The
Landau theory of the Fermi liquid in dimension $D > 1$ can be
taken as a paradigm of the success of this program. It has
been shown that, at least in the continuum limit, a system with
isotropic Fermi surface and regular interaction is susceptible 
of developping a fixed-point in which the interaction remains
stable in the infrared\cite{sh}. This example is of particular
significance since there are systems that 
do not lie in the perturbative regime, and yet we know that
their description in terms of the effective Fermi liquid theory is
correct, as its predictions apply fairly well.

The question of whether different critical points may arise at
dimension $D = 2$ is now a subject of debate\cite{a,feld,fk,wm}. 
The interest has
rosen in parallel to the failure of a global theoretical
understanding of the high-$T_c$ superconductivity of the
cuprates. The two-dimensional Hubbard model
has been proposed long time ago as the correct starting point to
study the electronic correlations inside the copper oxide
planes\cite{and}, and several authors have also proposed recently that it
should have a quantum critical point separating
antiferromagnetic and d-wave superconducting phases\cite{qcp}. However,
the elucidation of the existence of such critical point is quite
difficult, specially in the
strong coupling regime, since the presence of the strong
correlations do not make obvious even the determination of the
continuum limit of the model. Actually, apart from the above mentioned case
of regular Fermi surface, in general it is not possible to find 
by simple inspection of the microscopic hamiltonian
a set of scaling operators that behave properly under RG
transformations. This is particularly true in the case of the
Hubbard model at half-filling.

The above considerations stress the interest of identifying
microscopic models with correct scaling properties from the RG
point of view. The $t-t'$ Hubbard model filled up to the level
of the Van Hove singularity provides such an example. The
determination of the scaling properties and low-energy phases
for repulsive interaction has been accomplished in Ref.
\onlinecite{rgttp} (see also Ref. \onlinecite{mark}). 
We are interested now in the case of the $t-t'$
Hubbard model at negative $U$, which highlights some important
properties of the electron interactions near a Van Hove
singularity. 

We briefly review the way in which the low-energy limit 
is taken in the
model, and how in this case a set of scaling operators can be
read at once from the hamiltonian\cite{vanhove1}.
High-energy and low-energy electron modes are separated by
an energy cutoff $E_c$, that is sent progressively towards 
the Fermi line as high-energy modes are integrated out in
the RG process. 
When the Fermi level is at
the Van Hove singularity, as shown in Fig. \ref{one}, 
most part of the low-energy states
close to the Fermi line are concentrated around the saddle
points at $(\pi, 0)$ and $(0, \pi)$, as these features are at
the origin of the divergent density of states. Therefore, in
building up the low-energy effective theory we may focus on two
patches around the respective saddle points $A$ and $B$, where the
dispersion relation can be approximated by
\begin{equation}
\varepsilon_{A,B} ( {\bf k} ) \approx \mp ( t \mp 2 t' ) k_x^2 a^2
\pm ( t \pm 2 t' ) k_y^2 a^2
\end{equation}
$a$ being the lattice constant. In the continuum limit the rest
of modes are irrelevant, from the RG point of view. In fact, the
effective action for the low-energy modes restricted to the
region $|\varepsilon_{\alpha} \; ({\bf k})| \leq E_c $ is given
by 
\begin{eqnarray}
S  & = & \int d \omega d^2 k \sum_{\alpha,\sigma} \left( \omega
\; a^{+}_{\alpha,\sigma}({\bf k}, \omega )
   a_{\alpha,\sigma}({\bf k}, \omega )
- \varepsilon_{\alpha} ( {\bf k}) \;
   a^{+}_{\alpha,\sigma}({\bf k}, \omega )
   a_{\alpha,\sigma}({\bf k}, \omega ) \right)   \nonumber   \\
  &   &  - U \int d\omega d^2 k \;
 \rho_{\uparrow } ({\bf k}, \omega) \;
      \rho_{\downarrow } (-{\bf k}, -\omega)
\label{actk}
\end{eqnarray}
where $a_{\alpha,\sigma} (a^{+}_{\alpha,\sigma})$ are electron
annihilation (creation) operators ($\alpha$ labels the Van Hove point)
and $ \rho_{\uparrow ,\downarrow }$ are the density operators in
momentum space. Under a change in the cutoff $E_c \rightarrow s
E_c$, with a corresponding scaling of the momenta ${\bf k}
\rightarrow s^{1/2} {\bf k}$, one can check that the effective
action remains scale invariant after an appropriate scale
transformation of the electron modes $a_{\alpha,\sigma}
\rightarrow s^{-3/2} a_{\alpha,\sigma}$ \cite{vanhove1}. 
We have therefore a
microscopic model with a well-defined scaling behavior,
susceptible of being studied by means of RG methods.

As $E_c$ is sent towards the Fermi level and the electron modes
are labelled according to the saddle point they are attached to,
the interactions may be classified in the form shown 
in Fig. \ref{two}. 
The four possible types of local interaction $U_{intra}$,
$U_{inter}$, $U_{back}$ and $U_{umk}$ are renormalized by
quantum corrections, following a conventional pattern from the
quantum field theory point of view. By integrating high-energy
excitations in the slices $E_c - dE_c < |\varepsilon | < E_c$,
the lowest order $O(dE_c / E_c)$ corrections are given by the
particle-hole diagram of Fig. \ref{three}(a). 
It is worthwhile to stress
that, within this wilsonian RG approach, there are no more
diagrams renormalizing the interaction coupling constants, since
the particle-particle channel produces a contribution that
is in general $\sim (dE_c )^2$. This point has been conveniently
clarified in Ref. \onlinecite{sh}. It is only when the
total momentum of the colliding particles equals zero that the
particle-particle channel develops a contribution $\sim dE_c$, 
but this points at the correction of a correlation
function, defined at a particular momentum value, rather than at 
the renormalization of a coupling constant.

As remarked in Ref. \onlinecite{rgttp}, the diagram 
in Fig. \ref{three}(a)
corresponds to an {\em antiscreening} effect, that tends to
enhance repulsive interactions. By the same reason, its effect
in the present model is that of reducing the attractive
interaction and cannot produce by itself any instability driving
away from normal metallic behavior. The renormalized on-site 
interaction has the asymptotic behavior $U \sim 1/ \left| \log 
E_c \right| $ as the high-energy cutoff is reduced. 
As observed in Ref.
\onlinecite{vanhove2}, however, there are effective interactions
generated by quantum corrections that are not present in the
original hamiltonian. These are interactions between currents with
parallel spin, that appear through second order processes like
that shown in Fig. \ref{three}(b). They are strongly momentum dependent,
as the polarizabilities near ${\bf q} = 0$ and ${\bf q} = {\bf Q} 
\equiv (\pi, \pi )$ are given respectively by
\begin{eqnarray}
\chi ({\bf q}, \omega = 0 ) & = & \frac{c}{2\pi^2 t} \log \left|
\frac{E_c}{\varepsilon ({\bf q}) } \right|    \\
\chi ' ({\bf q}, \omega = 0 ) & = & \frac{c'}{2\pi^2 t} \log \left|
\frac{E_c}{ t a^2 ({\bf q} - {\bf Q})^2 } \right|
\end{eqnarray}
where  $c \equiv 1/\sqrt{1 - 4(t'/t)^2}$ and $c' \equiv
\log \left[ \left(1 + \sqrt{1 - 4(t'/t)^2} \right)/(2t'/t) \right]$.
These interactions give rise to a potential that is actually 
singular at small
momentum transfer. In real space this corresponds to an
interaction decaying like $\sim 1/r^2 $. It is not strange that
this kind of interaction may arise in the renormalization of the
model since, together with the purely local interaction, it
corresponds to the other not irrelevant four-fermion operator
that may appear in the effective action. According to the
wilsonian RG approach, once these effective interactions appear
they have to be considered on the same footing than the original
bare interactions, since they are needed for the complete
renormalization of the model. 

The processes shown in Fig. \ref{three}(b) are singular at small momentum
transfer and at momentum transfer $\sim {\bf Q} \equiv (\pi,
\pi)$. Therefore we have to introduce new couplings
$V_{intra}$, $V_{inter}$, $V_{back}$ and $V_{umk}$ for the
effective potential between electrons
with parallel spins, according to the classification in Fig. \ref{two}.
The important point is that these new interactions are also
attractive since they are due to {\em overscreening}, that is,
to the screening of an interaction vanishing at the classical
level.

The overscreening effect can be also understood
within the RG framework, by solving
the flow equations with the initial conditions
$V_{intra}(E_c)=V_{inter}(E_c)=
V_{back}(E_c)=V_{umk}(E_c)=0$  
at the upper value $E_c$ of the cutoff. The most significant
contributions to the renormalization of the couplings
are given by the diagram in Fig. \ref{three}(b) at small momentum
transfer, for $V_{intra}$ and $V_{inter}$, and at momentum
transfer ${\bf Q}$, for $V_{back}$ and $V_{umk}$. In the diagram
the interaction lines may stand either for the local interaction
potential or for the $1/r^2$ potential. One has to realize that,
under renormalization, $1/r^2$ effective interactions may also
arise between currents with opposite spin, that we denote,
conserving the previous notation, by $V_{\perp intra}$,
$V_{\perp inter}$, $V_{\perp back}$ and $V_{\perp umk}$. The
RG flow equations, that reflect the change of the couplings
by integration of particle-hole processes at the high-energy
cutoff $E_c$, are given by
\begin{eqnarray}
E_c \frac{\partial }{\partial E_c}\left( V_{intra} \pm V_{inter} \right)  
  & = &  -\frac{1}{2\pi^2 t} c \left( \left( U_{intra} \pm U_{inter} 
 \right)^2 + \left( V_{intra} \pm V_{inter}  \right)^2 +
 \left( V_{\perp intra} \pm V_{\perp inter} \right)^2 \right)  \\ 
E_c \frac{\partial }{\partial E_c}\left( V_{\perp intra} \pm
  V_{\perp inter} \right)  
  & = &  -\frac{1}{2\pi^2 t} c \left( 2 \left( U_{intra} \pm U_{inter} 
 \right) \left( V_{intra} \pm V_{inter} \right) +
                      2 \left( V_{\perp intra} \pm V_{\perp inter} 
 \right) \left( V_{intra} \pm V_{inter} \right) \right)
\end{eqnarray}
and two more equations obtained from the former two by the 
replacements $c \leftrightarrow c'$, $X_{intra} \leftrightarrow
X_{back}$ and $X_{inter} \leftrightarrow X_{umk}$ ($X = U, V,
V_{\perp}$). 
Given that at the begining of the RG process all the
interactions vanish, except $U_{intra}$, $U_{inter}$, $U_{back}$ 
and $U_{umk}$ that equal the value of the local attraction $U$,
it is easily seen that the instabilities of the model are given
by the flow of the $V$ and $V_{\perp}$ couplings, that become
increasingly attractive at low energies.

The instabilities of the coupling constants characterize the
different ground states of the system, which may be determined by
studying the different response functions. These are correlators
computed at particular values of the momentum, which signal the way
in which symmetry breakdown takes place in the model. One may
check that in the present case of attractive interaction ($U <
0$), the only correlators that may diverge at a given frequency
are those for the operators $\sum_{{\bf k}} \left(
a^{+}_{A\uparrow }({\bf k}) a^{+}_{A\downarrow }({\bf -k}) +
a^{+}_{B\uparrow }({\bf k}) a^{+}_{B\downarrow }({\bf -k}) +
 h.c. \right)$, $\rho_{\uparrow} ({\bf Q},\omega ) +
\rho_{\downarrow} ({\bf Q},\omega ) $ and
$\rho_{\uparrow} ({\bf 0},\omega ) +
\rho_{\downarrow} ({\bf 0},\omega ) $. Their response functions
characterize, respectively, s-wave superconductivity ($R_{SCs}$),
charge density wave ($R_{CDW}$) and phase separation ($R_{PS}$)
instabilities.

We compute the response functions by exploiting again the scaling
properties of the model. For this purpose we establish the
dependence of the correlators on the cutoff $E_c$ and, taking into
account the scale invariance of the model (up to logarithmic
renormalizations of the couplings), we introduce the dimensionless
scaling variable $E_c / \omega $  to determine the dependence on
the frequency $\omega$. The starting point is the perturbative
computation of each response function, that exhibits a logarithmic
dependence on $E_c / \omega $. In contrast to the RPA, we do not
sum up the iteration of particle-hole bubbles, as this cannot
be a reliable expansion of the correlator in a model with strong
renormalization of the one-particle properties\cite{vanhove2}.
Instead we compute
the variation of the correlator under a reduction of $E_c$ and
write it down in terms of the correlator and coupling constants at
the new value of the cutoff. This procedure is similar to that
followed in the study of one-dimensional electron systems\cite{1d}
or coupled one-dimensional chains\cite{bf}.

The
scaling equations in the present case turn out to be
\begin{eqnarray}
\frac{\partial R_{CDW}}{\partial E_c} & = & -  \frac{2c'}{\pi^2 t}
  \frac{1}{E_c}  -  \frac{c'}{ \pi^2 t} \left( V_{back} +
  V_{umk} + V_{\perp back} + V_{\perp umk} \right) 
     \frac{1}{E_c} R_{CDW}     \label{rcdw}               \\
\frac{\partial R_{SCs}}{\partial E_c} & = & -  \frac{c}{2\pi^2 t}
  \frac{\log (E_c/\omega) }{E_c}  -
  \frac{c}{2 \pi^2 t} \left( V_{\perp intra} +
  V_{\perp umk} \right) \frac{\log (E_c/\omega) }{E_c} R_{SCs}  \\
\frac{\partial R_{PS}}{\partial E_c} & = & -  \frac{2c}{\pi^2 t}
  \frac{1}{E_c}  -  \frac{c}{ \pi^2 t} \left( V_{intra} +
  V_{inter} + V_{\perp intra} + V_{\perp inter} \right) 
     \frac{1}{E_c} R_{PS}     \label{rps}
\end{eqnarray}
The equation for $R_{SCs}$ shows the nontrivial scaling factor
$\log (E_c/\omega) $, as a consequence of the enhanced
susceptibility at zero momentum
in the particle-particle channel, that behaves as
$ \sim \log^2 (E_c/\omega) $. Such factor is nothing but a
reflection of the divergent density of states at
the Van Hove singularity, and
it does not spoil, in any event, the scaling properties of the
model. 

As remarked before, the natural scaling variable in the model is
$E_c / \omega $, what becomes now evident by inspection
of the scaling equations (\ref{rcdw})-(\ref{rps}). The fact
that the renormalized coupling constants follow an unbounded
flow may cast doubts in the reliability of the perturbative
RG approach. The divergences that, as a consequence of it, are
found in the response functions at certain values of $E_c /
\omega $ provide, however, a way of discerning the competition
between the different instabilities of the system.
The response function having the strongest divergence
as the value of $\omega /E_c $ is lowered characterizes the
ground state of the system. According to this criterion, we have
determined the different phases at weak $U$ coupling and $t'$ up
to $0.5t$, where the Fermi sea degenerates into a pair of
straight lines. The phase diagram is shown in Fig. \ref{four} .
We recall that the charge-density-wave and superconducting 
ground states are degenerated in the $t' = 0$ attractive 
Hubbard model at half-filling\cite{robas,auer}.
On the other hand, it has been shown that phase
separation cannot take place in the Hubbard model on a bipartite
lattice, at any value of the interaction\cite{su}. The proof of
this statement does not follow in our case, however, as the
next-to-nearest neighbor hopping $t'$ spoils the splitting into
two sublattices. 

Our results should be applicable in
the weak coupling regime and to Fermi surfaces corresponding to
an appreciable value of $t'$. This is because the main
assumption of our RG approach is the possibility of taking the
continuum limit, which neglects correlation effects due to a large
on-site attraction $U$. These are important for the Hubbard
model at or near half-filling. Otherwise, our approach
provides the evidence in a concrete model of the proposed
connection between charge-density-wave or phase separation
instabilites and anisotropic superconductivity\cite{perali}.
We have to bear
in mind that, in our low-energy theory, the condensate
wavefunction turns out to have the same sign near $(\pi, 0)$ and
$(0, \pi)$  but it must have necessarily less strength far from
the saddle points, leading to a highly anisotropic gap.
Specially sensible is our prediction of
phase separation at low filling and weak coupling, when the Fermi sea
is degenerating towards a pair of straight lines near $t' = 0.5 t$.
The interplay between phase separation and anisotropic
superconductivity has been also considered in the $t-J$ and U-V models 
\cite{dag} as well as in the three-band Hubbard model\cite{grilli}.
Phase separation near a Van Hove singularity has been discussed
in a model with electron-phonon interaction in Ref.
\onlinecite{mark2}. 

With regard to real systems,
the realization of a charge-density-wave in the model may bare
a direct relation with the recent experimental observation of
such instability in two-dimensional interfaces\cite{exp}.
In those systems the atoms at the surface are arranged in a
triangular lattice, but the most important point is the present
recognition that there is no significant nesting of the Fermi
line accounting for the instability. The form of the dispersion
relation has been evaluated in several instances, and it shows
that the Fermi line is close to the saddle points at the boundary
of the Brillouin Zone\cite{tossati}.

From the theoretical point of view,
one of the most remarkable features of the phase diagram in Fig.
\ref{four} 
is the existence of a triple point where the three
different phases coalesce. The phase diagram is actually a map
of the different ground states of the system. If we look at
phase separation as an instability in which no symmetry
breakdown takes place in a microscopic scale, opposite to what
happens in the other two phases, we may conjecture that the
triple point must correspond to a quantum critical point of the
system. In fact, it has to be possible to crossover smoothly
from the superconducting ground state to the charge-density-wave
state through the phase separation region. The crossing of the
two ground states is the distinctive feature of the quantum
phase transtition.

It is very suggestive the similarity that the phase diagram
of Fig. \ref{four} bears with that of the repulsive $t-t'$
Hubbard model at the Van Hove filling obtained with analogous
RG techniques\cite{rgttp}. In that case there is a large
boundary in the phase space between antiferromagnetic and
d-wave superconducting phases, up to a point where the leading
instability turns out to be ferromagnetism. The existence of
these three phases has been confirmed by numerical methods.
Before the RG analysis had been undertaken, quantum Monte Carlo
computations had already provided a clear signature of
ferromagnetism at $t' = 0.47 t$\cite{ferro}. On the other hand,
recent quantum Monte Carlo computations\cite{qmc}, as well as
mean-field computations with effective interactions\cite{mf},
are supporting the existence of a d-wave superconducting phase
at intermediate values of $t'$. These evidences prove the
predictability of the RG approach to the present model, and they
reassure us that the phases discussed above should be
susceptible of being obtained by alternative numerical
methods.
The question of the mentioned quantum critical
point is important enough for being tested by other
computational techniques.
A number of points, like the nature of the strong
coupling phases or the thermodynamic properties of the
model, should be also investigated to that effect.

It is a pleasure to thank fruitful discussions with F. Guinea,
S. Sorella and M. A. H. Vozmediano. This work has been
partially supported by the CICYT Grant  PB96-0875.

\begin{figure}
\caption{Energy contour lines about the Fermi level, with the
Fermi line passing by the saddle points A and B.}
\label{one}
\end{figure}

\begin{figure}
\caption{Different interaction terms arising from the flavor
indexes A and B.}
\label{two}
\end{figure}

\begin{figure}
\caption{Second order diagrams renormalizing the
interactions in the model, with electron lines carrying flavor
index A or B appropriate to each case.}
\label{three}
\end{figure}

\begin{figure}
\caption{Phase diagram in the $(t',U)$ plane showing the regions
of charge-density-wave (CDW), s-wave superconductivity (SC-s) and
phase separation (PS) instability.}
\label{four}
\end{figure}

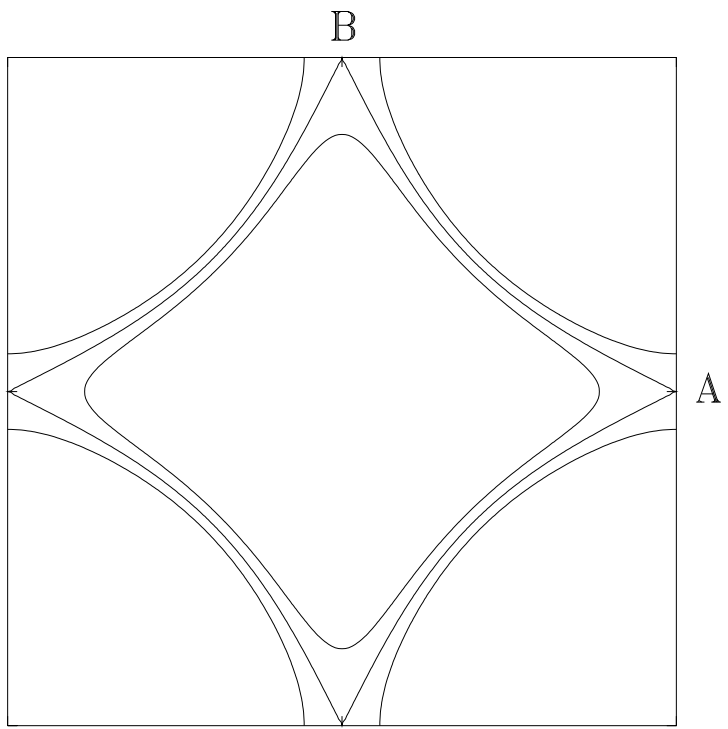
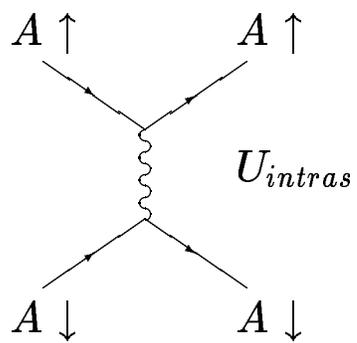
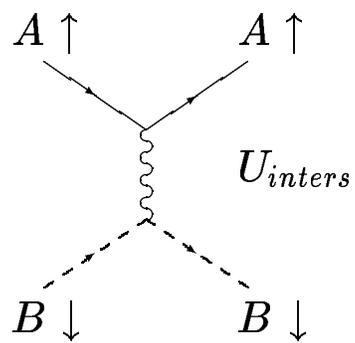
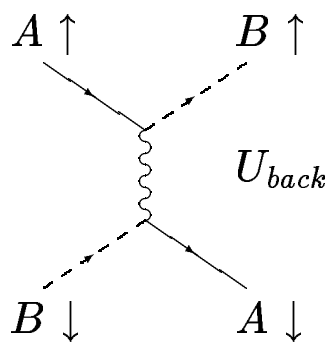
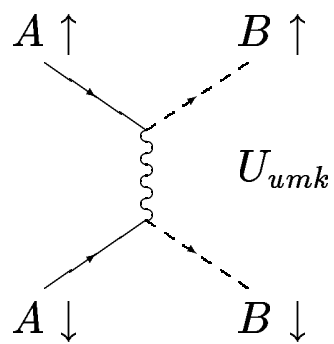
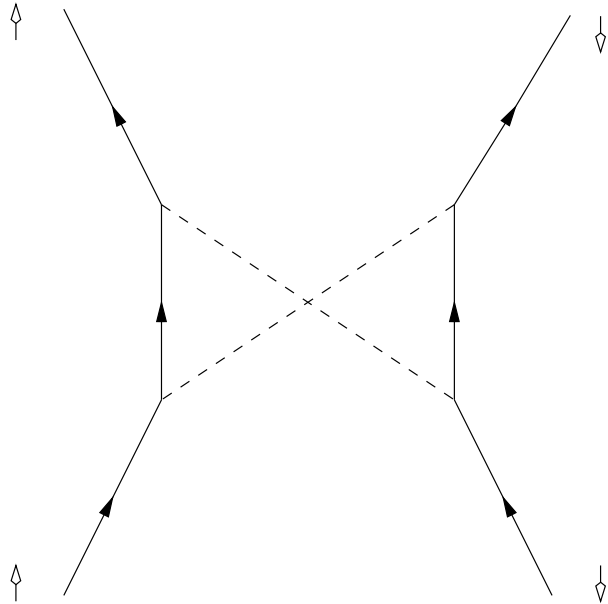
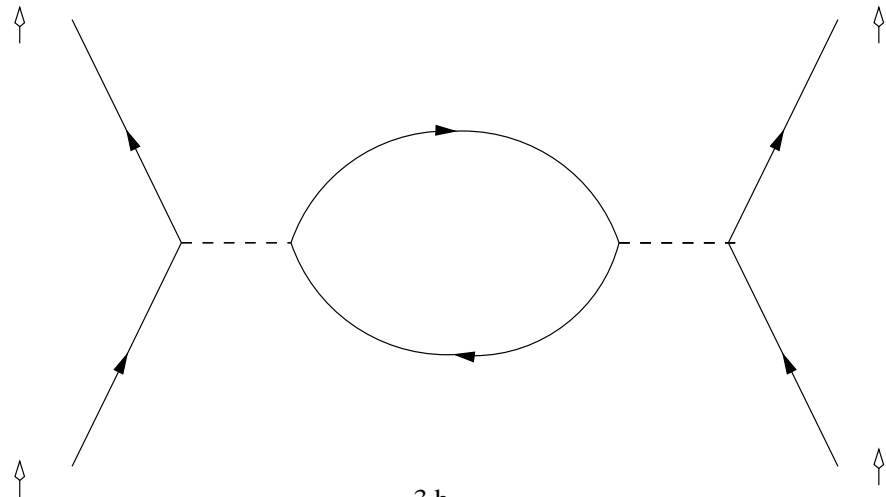
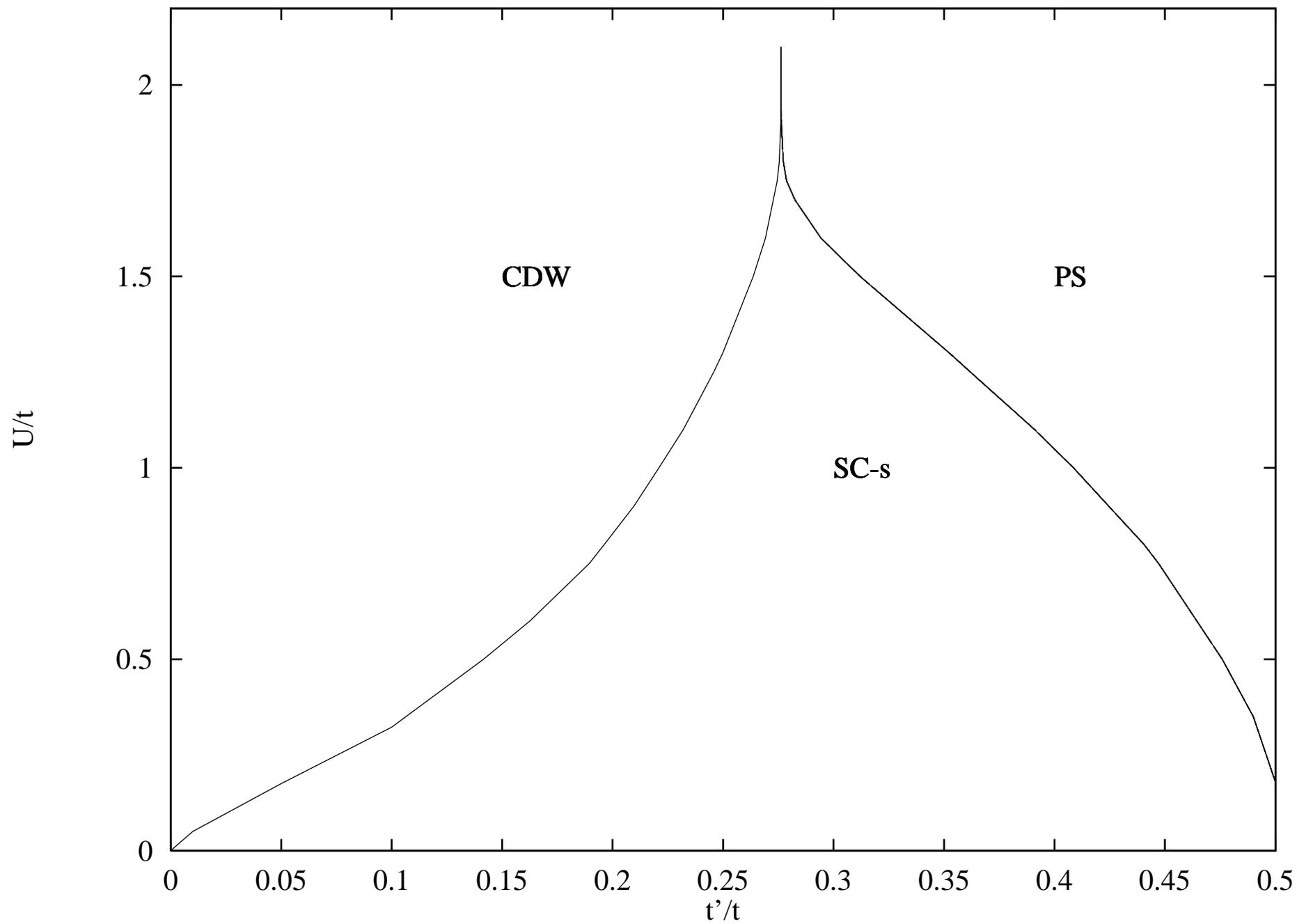
\end{document}